\documentclass[prl,twocolumn]{revtex4-1}

\usepackage[T1]{fontenc}
\usepackage[utf8]{inputenc}
\usepackage{amsmath}
\usepackage{amssymb}
\usepackage{graphicx}
\usepackage{float}
\usepackage{bm}
\usepackage[colorlinks]{hyperref}
\usepackage{xcolor}
\usepackage{adjustbox}
\makeatletter

\newcommand{\beginsupplement}{%
        \setcounter{table}{0}
        \renewcommand{\thetable}{S\arabic{table}}%
        \setcounter{figure}{0}
        \renewcommand{\thefigure}{S\arabic{figure}}%
        \setcounter{section}{0}
        \renewcommand{\thesection}{S\arabic{section}}%
        \setcounter{section}{0}
        \renewcommand{\thesection}{S\arabic{section}}%
        \setcounter{subsection}{0}
        \renewcommand{\thesubsection}{S\arabic{section}.\arabic{subsection}}%
        \setcounter{equation}{0}
        \renewcommand{\theequation}{S\arabic{equation}}%
     }
     
\makeatother

\global\long\def\ket#1{\left| #1\right\rangle }
\global\long\def\bra#1{\left\langle #1 \right|}

\global\long\def\av#1{\left\langle #1 \right\rangle }
\global\long\def\tr{\text{tr}}
\global\long\def\Tr{\text{Tr}}

\definecolor{applegreen}{rgb}{0.55, 0.71, 0.0}

\begin{document}

\title{Efficient quantum information probes of non-equilibrium quantum criticality}

\author{Miguel M. Oliveira}
\email{miguel.m.oliveira@tecnico.ulisboa.pt}
\affiliation{CeFEMA, Instituto Superior T\'ecnico, Universidade de Lisboa Av. Rovisco
Pais, 1049-001 Lisboa, Portugal}

\author{Pedro Ribeiro}
\email{ribeiro.pedro@tecnico.ulisboa.pt}
\affiliation{CeFEMA, Instituto Superior T\'ecnico, Universidade de Lisboa Av. Rovisco
Pais, 1049-001 Lisboa, Portugal}
\affiliation{Beijing Computational Science Research Center, Beijing 100193, China}

\author{Stefan Kirchner}
\email{stefan.kirchner@correlated-matter.com}
\affiliation{Department of Electrophysics, National Yang Ming Chiao Tung University, Hsinchu 30010, Taiwan}
\affiliation{Center for Emergent Functional Matter Science, National Yang Ming Chiao Tung University, Hsinchu 30010, Taiwan}

\begin{abstract}
Quantum information-based approaches, in particular the fidelity, have been  flexible probes for phase boundaries of quantum matter. 
A major hurdle to a more widespread application of fidelity and other quantum information measures to strongly correlated quantum materials is the inaccessibility of the fidelity susceptibility to most state-of-the-art numerical methods. 
This is particularly apparent away from equilibrium where, at present, no general critical theory is available and many standard techniques fail.
Motivated by the usefulness of quantum information based measures we show that a widely accessible quantity, the single-particle affinity, is able to serve as a versatile instrument to identify phase transitions beyond Landau's paradigm.
We demonstrate that it not only is able to signal previously identified  non-equilibrium phase transitions but also has the potential to  detect hitherto unknown phases in models of quantum matter far from equilibrium.
\end{abstract}
\maketitle

\section*{Introduction}

Quantum fluctuations drive zero temperature quantum phase transitions\cite{sachdev2011}. In equilibrium, the fluctuation-induced long-range entanglement has been successfully used as a probe of quantum criticality\cite{Osterloh2002,SHI-JIAN}. 

Away from equilibrium the situation seems less clear. Quantum  systems far from equilibrium can exhibit a wide variety of phenomena ranging from the usual symmetry-based phase transitions \cite{Morrison2008,Honing2012,Lee2014,Genway2014,Manzano2014,Wilming2017,Sanchez2019,Ferreira1,Julian2021,Kessler_2012} to the more exotic behavior, such as mixed-order phase transitions \cite{Tharnier,Larson2018,Julian2020} or time crystallinity \cite{Khemani2016,Else2016}, with no equilibrium counterpart.
Probing such systems necessarily requires an understanding of their non-equilibrium properties. While a general theory of far-from equilibrium criticality is not available, it is known that not all systems are describable in terms of  order parameter fluctuations \cite{Ribeiro2015}. 
Already the mere detection of a phase transition away from equilibrium, where thermodynamic minimization principles no longer apply, can be challenging, in particular if the  order parameter is elusive. 
Likewise, detecting phase transitions based on spectral properties require information on excited states that is also hard to access by most methods.

Quantum information-centered approaches, based on the fidelity, are natural candidates as 'all purpose' quantities to circumvent that issue, i.e. to detect phase transitions in the absence of additional knowledge\cite{Zanardi2007,Zanardi2007-2,Cozzini2007,Rossini2018,Mera2018}. However, for interacting  many-body systems its computation requires information about the full density matrix. This makes it forbiddingly difficult, if not impossible, to compute the fidelity as most state of the art numerical techniques only allow for the faithful determination of few-body observables. This renders the fidelity inaccessible to Quantum Monte Carlo methods and rescinds the advantage of the exponential compression exploited by variational methods such as the density matrix renormalization group (DMRG) for mixed states.

Single-particle correlators, which form the basic building block of many-body theory, are in contrast readily accessible.
In equilibrium, the use of the single-particle correlator matrix was first proposed in the context of the superfluidity transition \cite{Penrose}. Similar methods have recently been employed to test for many-body localization in closed systems \cite{Bera1,Bera2}.
Thus, the idea of extending quantum information concepts, in particular the fidelity, to single-particle observables as a versatile tool  to detect phase transitions away from equilibrium presents itself.   

Here, we provide a proof-of-concept of this idea by establishing that a class of single-particle observables, referred to as single-particle distances, can detect phase transitions out of equilibrium. These quantities are derived from a notion of proximity between quantum states. 
Being dependent solely on single-particle quantities, these distances can be efficiently evaluated by commonly employed numerical methods. We demonstrate that they can be used in the detection of phase transitions between mixed states, arising  in open quantum system setups and also in equilibrium at finite temperatures. This observation is particularly pertinent in the out-of-equilibrium case due to a lack of alternative methods.
In what follows, we focus on the single-particle affinity (SPA) defined below, which exactly reduces to the fidelity for Gaussian states. 

We illustrate the usefulness of the SPA through the discussion of a model with a well-established non-equilibrium steady-state (NESS) phase diagram. We then apply the SPA as an 'all purpose' detector of phase transitions using it to investigate a boundary-driven fermionic ladder whose NESS phase diagram has not been reported so far. As it turns out, this phase diagram is rather rich, contrary to  naive expectations. This thus establishes the usefulness of the SPA.
We also confirm the findings based on the SPA through a careful analysis of the finite size scaling of the current.

\section*{Results}

\subsection*{Single-particle affinity}

In order to assess the far-from-equilibrium behavior of electronic matter, we note that for a fermionic system all single-particle observables can be obtained from the covariance matrix  $\bm{\Sigma}= \langle \mathbf{C} \, \mathbf{C}^\dagger \rangle$, where $\mathbf{C}=\left( c_1, c_2, \cdots, c_1^\dagger, c_2^\dagger, \cdots \right)^T$ is a  Nambu-vector. Here, $\bm{\Sigma}$ is Hermitian, $\bm{\Sigma}^\dagger=\bm{\Sigma}$ and respects charge conjugation symmetry, i.e., $\bm{\Sigma} = \tau_x \left( 1-\bm{\Sigma}^T \right) \tau_x $, with $\tau_x$ being the $x$ Pauli matrix acting in Nambu space. 

It is well known that in equilibrium, the order parameter of, e.g., charge-, spin-density waves or superconducting phases is constructed from single-particle operators, which can even signal topological phase transitions for which a local order parameter cannot be defined. 
Even far from equilibrium, where a symmetry-breaking order parameter is often not available, the covariance matrix may still  encode important information about phase changes in the NESS.  

We explore the ability of the covariance matrix to identify phase transitions by studying a suitable measure of distance between covariance matrices  $D_{\mathcal A} (\bm{\Sigma}_1,\bm{\Sigma}_2) = \sqrt{2-2\sqrt{ \mathcal A (\bm{\Sigma}_1,\bm{\Sigma}_2) }}$ where $\mathcal A $ is a quantity we will refer to as SPA. It is defined by
\begin{equation} 
\mathcal A (\bm{\Sigma}_1,\bm{\Sigma}_2)=  \frac{ \det \left[ \mathbf{1} + \sqrt{\left( \bm{\Sigma}_1^{-1} - \mathbf{1} \right) \left( \bm{\Sigma}_2^{-1} - \mathbf{1} \right) } \right] }{\sqrt{\det \left[ \bm{\Sigma}_1^{-1} \right] \det \left[ \bm{\Sigma}_2^{-1} \right] }}. 
\label{FidX1}
\end{equation}
$D_{\mathcal A}$ possesses all the properties of a metric, thus providing a sensible notion of affinity between two states (see the Methods section). 
For non-interacting systems the density matrices are Gaussian, in which case $\mathcal A$ coincides with the fidelity $F$, i.e., $F(\rho_1,\rho_2)=\mathcal A(\bm{\Sigma}_1,\bm{\Sigma}_2)$, where $F(\rho_1,\rho_2) = \left(\Tr \, \sqrt{\sqrt{\rho_1} \, \rho_2 \, \sqrt{\rho_1}} \right)^2$ and $\bm{\Sigma}_i= \Tr \left[ \mathbf{C} \, \mathbf{C}^\dagger \rho_ i\right]$. Therefore, in the quadratic case $D_{\mathcal A}(\bm{\Sigma}_1,\bm{\Sigma}_2)$ reduces to the Bures distance $D_{\mathcal B}(\rho_1,\rho_2) = \sqrt{2 - 2\sqrt{F (\rho_1,\rho_2) }}$ \cite{Bures69,Uhlmann76,Jozsa94,Uhlmann95}.

As the quantities entering $\bm\Sigma$ are, at least in principle, straightforwardly accessible in most numerical and approximation methods, detecting phase-transitions based on this quantity is of great practical relevance. This is in contrast to the fidelity whose numerical calculation, for an interacting system, is in general not feasible.  
To address this issue, we demonstrate the usefulness of $\mathcal A$ for detecting  steady-state phase transitions of interacting systems which can be evaluated efficiently. This is accomplished by studying a generalized linear response susceptibility associated with $\mathcal A$ and defined through
\begin{equation} 
\mathcal A [\bm{\Sigma}(\boldsymbol \lambda ) ,\bm{\Sigma}(\boldsymbol \lambda + \boldsymbol{ d\lambda} )] = 1- \sum_{i,j} \chi^{i,j}_\mathcal{A}(\boldsymbol \lambda ) \frac{d\lambda_i \, d\lambda_j}{2} + O(d \lambda^3) ,
\label{FidX}
\end{equation}
where $\boldsymbol \lambda $ parametrizes the NESS. Here, the first derivative term is absent since the expansion is done around the maximum. By construction, $\mathcal A$ reproduces previous results for non-interacting fermionic systems based on $F$ \cite{Zanardi2007-2} and $ \chi^{i,j}_\mathcal{A}$ reduces to the fidelity susceptibility.

\subsection*{Application to boundary driven systems}

%%%%%%%%%%%%%%%%%%%%%%% Figure 1 %%%%%%%%%%%%%%%%%%%%%%%%%%
\begin{figure}[t!]
\centering
\includegraphics[width= 0.4 \textwidth]{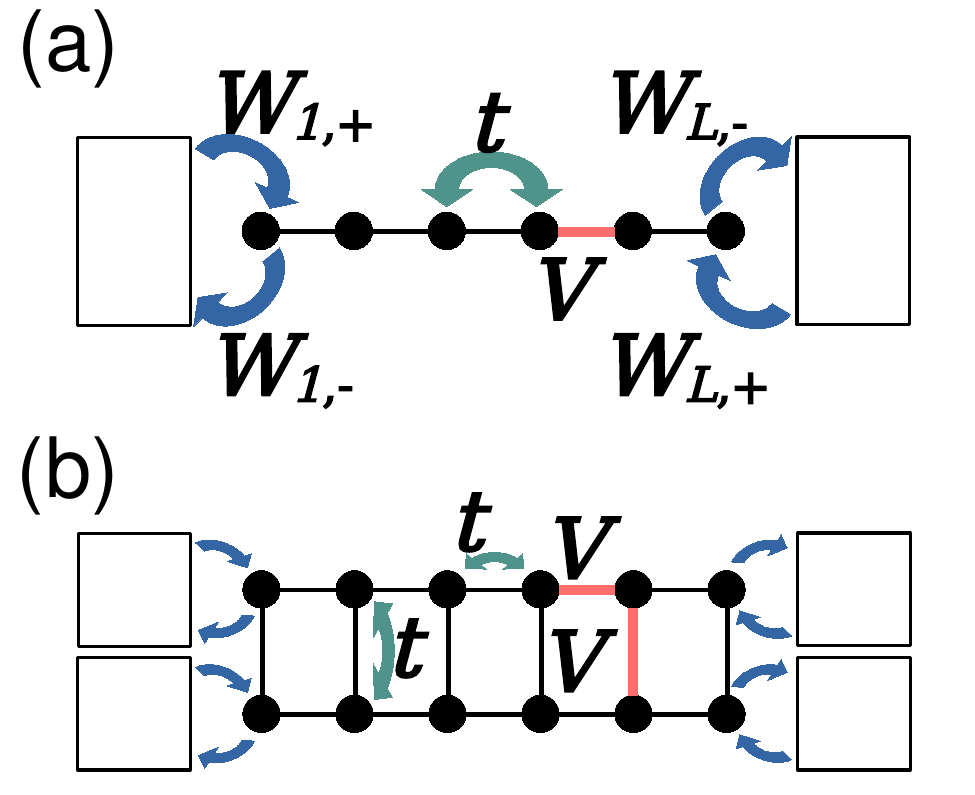} 
\caption{\textbf{Schematics of the studied models.} Fermion chain (a) and fermion ladder (b).}
\label{sketches}
\end{figure}
%%%%%%%%%%%%%%%%%%%%%%%%%%%%%%%%%%%%%%%%%%%%%%%%%%%%%%%%%%

The methodology proposed here is applicable to open quantum systems. An important subset of those are systems in the so-called Markovian regime on which we focus in what follows. 
Within the Markovian approximation time scales of the environment are taken to be much shorter than those of the system. This Markovian limit has recently received considerable attention \cite{landi2021nonequilibrium} both for its physical relevance and because it represents a substantial simplification with respect to generic open many-body quantum systems. 
Additionally, for a wide class of one dimensional models, non-equilibrium steady-states (NESS) of Markovian systems can be effectively parameterized by matrix product operators (MPO) \cite{Vidal3,Cirar1,Rams_2020,Gullans_2019,Cubitt_2015,Brandao_2015}. 
This approach leads to a number of important developments for the transport properties of quantum systems and in particular spin chains.
For a boundary driven Heisenberg XXZ chain, e.g., it helped establish the NESS phase diagram \cite{Prosen1,Benenti1,Benenti2,Marko1,Prosen2,Marko4,Marko5,Mendoza1,Mendoza2}. Further  support came from  a series of exact results \cite{Marko2,Marko3,Prosen3,Prosen4,Prosen7}.

We consider two boundary-driven models with Markovian reservoirs that allow injection or removal of electrons.
The Markovian evolution is described by a Lindblad equation  \cite{book1,book2}
\begin{equation}
\partial_t \rho= \mathcal{L}(\rho),
\label{Lind}
\end{equation}
where $\rho$ is the density matrix of the system. The Lindblad operator $\mathcal{L}$ is given by
\begin{equation}
\mathcal{L}(\rho) = -i [H,\rho] + \sum_\alpha \left( W_\alpha^\dagger \, \rho \, W_\alpha -\frac{1}{2} \left\{W_\alpha^\dagger W_\alpha, \rho \right\}  \right). \label{Lindblad}
\end{equation}
Here $[ \ ]$ and $\{\}$ denote commutator and anti-commutator respectively, while $W_l$ are the so-called jump operators which encode the  system-reservoir couplings. In Eq.\ (\ref{Lindblad}), the first term is responsible for the unitary part of the time evolution and the second describes driving and dissipation.

We analyse the steady-state properties of the models defined below using techniques for open systems based on MPS, which have been shown to yield reliable results for this class of boundary-driven problems \cite{Vidal3,Cirar1}. 
Starting in the infinite temperature state, we time evolve the system according to Eq.\eqref{Lindblad} using the t-DMRG algorithm \cite{Daley} until it reaches the steady state. Details of the implementation and convergence of the algorithm are provided in the Methods section and the supplementary material--S4. Except when explicitly stated otherwise, all numerical results  were obtained using matrix product state (MPS) techniques \cite{Vidal3,Cirar1,Prosen1}.

\subsection*{Fermionic chain}

The first system we consider  is the $t-V$ model for spinless electrons on a chain. We demonstrate that $\mathcal A$ reproduces the previously known phase diagram.  
A sketch of the model is provided in Fig.\ref{sketches}-(a) and
its Hamiltonian is given by (see also \eqref{H_chain} of Method section) 
\begin{equation}
H=  \sum_{\left\langle i,j \right\rangle } \left[-t\,c^\dagger_i \, c_{j}  + \frac{V}{2}  \left ( n_i - \frac{1}{2} \right)\left ( n_{j} - \frac{1}{2} \right) \right] ,
\label{Ham}
\end{equation}
where $c^\dagger_i, c_i$ are the creation and annihilation operators on site $i$, $n_i=c^\dagger_{i} \, c_{i}$, $t$ is the hopping amplitude, set to unity in the following, $V$ is a nearest-neighbour density-density interaction and $L$ the length of the system. Here, $i= 1,..., L$, and the nearest neighbour summation, $\sum_{\left\langle i,j \right\rangle }$, is restricted to $j = i\pm 1$, where open boundary conditions are assumed.
The jump operators in this case are 
\begin{align}
W_{l,-} =\sqrt{\Gamma_{l}\frac{1-\eta_{l}}{2}}\, c_{l}, \ \ 
W_{l,+} =\sqrt{\Gamma_{l}\frac{1+\eta_{l}}{2}}\, c_{l}^\dagger,
\label{Jump}
\end{align}
where $l=1,L$ labels the end points of the chain, $\Gamma_{l}$ is the injection/removal rate for the $l^{\text{th}}$ reservoir and the bias $\eta_{l}$ specifies the associated imbalance between particle injection and removal. Thus, the summation in Eq.\eqref{Lindblad} is performed over $\alpha=(l,\pm) $. In what follows we set $\Gamma_l=1$ and reduce the bias to a single parameter $\eta_1=-\eta_L=\eta$.
As a result, Eq.~(\ref{Ham}) is equivalent to the boundary-driven XXZ chain studied in Refs. \cite{Prosen1,Benenti1,Benenti2,Marko1,Prosen2,Marko4,Marko5,Mendoza1,Mendoza2,Prosen3,Prosen4}.

%%%%%%%%%%%%%%%%%%%%%%% Figure 2 %%%%%%%%%%%%%%%%%%%%%%%%%%
\begin{figure*}
\centering
\includegraphics[width= 1.0 \textwidth]{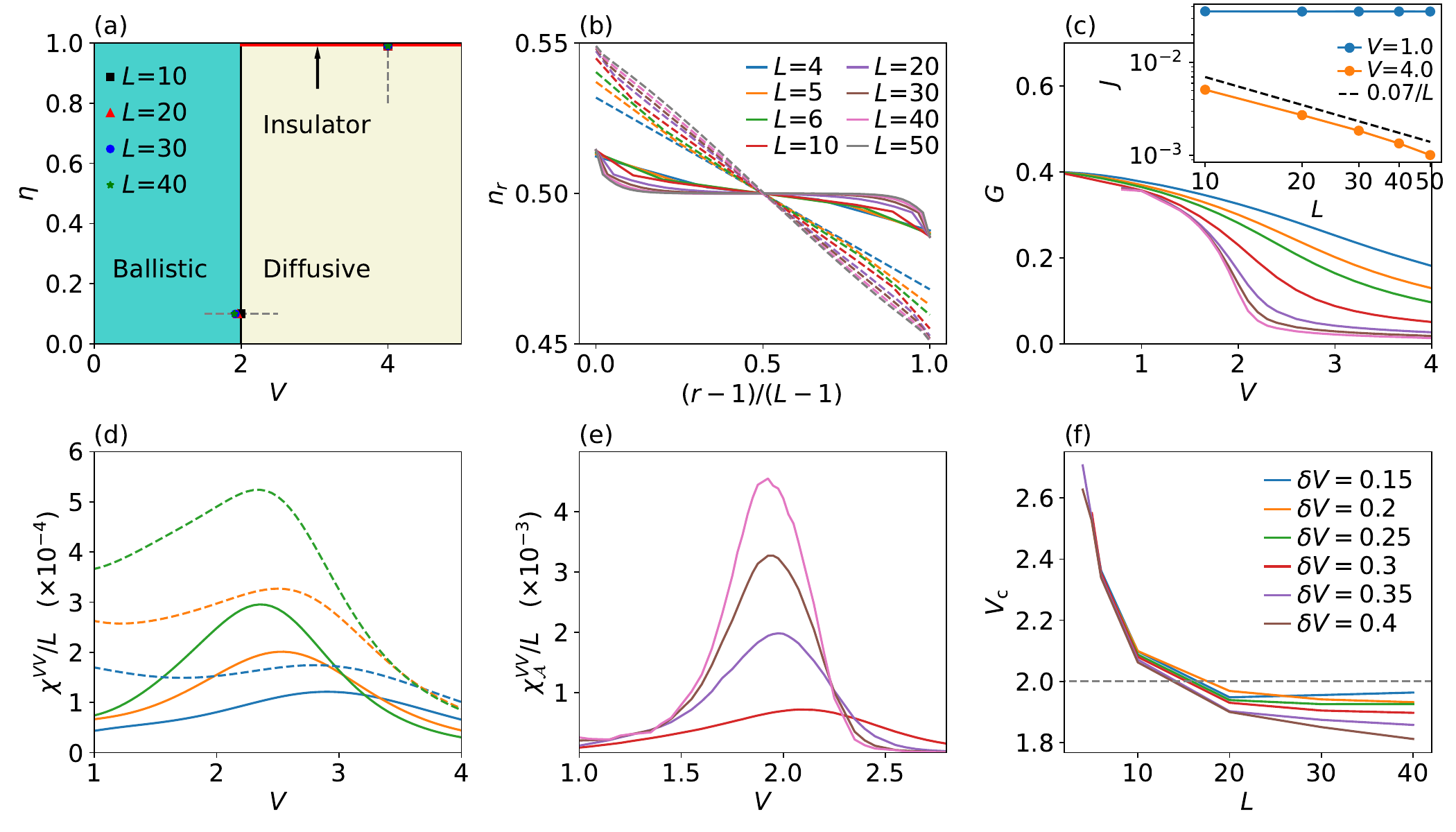}
\caption{\textbf{Results for the fermionic chain.} (a) Sketch of the known phase diagram, showcasing a ballistic, diffusive and insulating regimes. The symbols give the location of the peaks in (e) for different system sizes. (b) Comparison of the real space occupations between the ballistic phase at $V=1.0$ (continuous) and the diffusive at $V=4.0$ (dashed). 
Figures (b)-(f) are obtained for $\eta=0.1$. Figures (b)-(e) use the color code given in (b).
(c) $G$ versus $V$ for different system sizes. Inset shows the current's dependence with $L$ for the different regimes, with the dashed line being a guide for the eyes delineating the diffusive behavior. 
(d) Comparison of affinity (continuous) and fidelity (dashed) susceptibilities per degree of freedom obtained by exact diagonalization for $\delta V=0.01$. (e) $\chi_{\mathcal{A}}^{VV}/L$ for $\delta V=0.2$ obtained via MPS.  
(f) Critical coupling $V_\text{c}$, measured as maximum of $\chi^{VV}_{\mathcal{A}}$ vs $L$ for different perturbation sizes. 
}
\label{chain_pd}
\end{figure*}
%%%%%%%%%%%%%%%%%%%%%%%%%%%%%%%%%%%%%%%%%%%%%%%%%%%%%%%%%%

The steady-state phase diagram of the boundary-driven $t-V$ chain model is reproduced in Fig.\ref{chain_pd}-(a). 
Fig.\ref{chain_pd}-(b) shows the real space occupations, $n_r = c^\dagger_r c_r$. 
In the diffusive regime, away from the boundaries, the density profile is linear in $r$, i.e., 
%$n_r \simeq D(\eta) \eta  (r/L-1/2) + \bar{n} $, 
$n_r - 1/2 \propto  \eta (1/2-r/L) $.  This is in contrast to the ballistic case where the density profile is position independent away from the leads \cite{Prosen1,Benenti1,Benenti2}. 
The diffusion constant is then defined as
$J= - D(V,\eta) \frac{\partial n_r}{\partial r}, $
and becomes independent of $\eta$ for small $\eta$.

Fig.\ref{chain_pd}-(c) (inset) depicts the dependence of the current $J(L) = -i t \left\langle  c^\dagger_{i} \, c_{i+1} - \text{h.c.}  \right\rangle$ on $L$ in each of the phases: the ballistic phase, for $V<V_c=2$ , where  $J(L)$ is $L$ independent; the diffusive regime where $J(L) \propto L^{-1} $; and the insulating phase (not shown) for $\eta=1$ and $V>V_\text{c}$ where $J(L)$ vanishes exponentially with $L$ \cite{Benenti2}.

This leads us to the identification of a suitable order parameter to distinguish the possible regimes. We make use of the change of behavior of the current --  $L$-independent in the ballistic and  $\sim 1/L$ in the diffusive regime -- to introduce the conductance $G=J/\eta$, where $\eta$ takes the place of the applied bias. This quantity is shown in Fig.\ref{chain_pd}-(c) (main panel).

Given both the relative simplicity and the existence of many reliable results make this model an ideal benchmark for our method. 
Figs.\ref{chain_pd}-(d), (e) show the SPA susceptibility, $\chi^{VV}_{\mathcal{A}}\simeq -  \delta V^{-2} [ \mathcal{A}(V+ \delta V)-2 \mathcal{A}(V)+\mathcal{A}(V -\delta V)  ]$ (see Eq.\ref{FidX}), for the well established ballistic-diffusive transition of the chain model.  
Fig.\ref{chain_pd}-(d) also depicts a comparison with the fidelity susceptibility, $\chi^{VV}_{F}$, for small systems sizes, obtained by exact diagonalization. Clearly $\chi^{VV}_{F}$ signals the presence of a transition of the infinite system, at $V_\text{c}$, already for reduced system sizes and $\chi^{VV}_{\mathcal{A}}$ tracks the behavior of $\chi^{VV}_{F}$ near the transition.

Fig.\ref{chain_pd}-(f) shows the position of the maximum of $\chi_{\mathcal{A}}^{V V} $ as a function of $L$ for different perturbation sizes $\delta V$.
The finite-size scaling analysis confirms that $\chi^{VV}_{\mathcal{A}}$ can be used to effectively detect the steady-state phase transition located at $V_\text{c}=2$. 
The finite-size scaling analysis for fixed $\delta V$ leads to a $V_\text{c}(\delta V)<V_\text{c}$ which approaches $V_\text{c}$ as $\delta V$ decreases, i.e., $V_\text{c}(\delta V\rightarrow 0)\rightarrow V_\text{c}$. %shows a left-biased if $\delta V$ is not sufficiently small.
In practice, a decrease in $\delta V$ needs to be balanced against the concomitantly increasing computational effort.

We checked that the critical phase boundary in Fig.\ref{chain_pd}-(a) at $\eta=1$ is also detected by the proper affinity susceptibility, $\chi^{\eta\eta}_{\mathcal{A}}$,  both for the critical insulating phase ($V>V_\text{c}$) and for the critical ballistic phase ($V<V_\text{c}$) (see supplementary material--S2).

%%%%%%%%%%%%%%%%%%%%%%% Figure 3 %%%%%%%%%%%%%%%%%%%%%%%%%
\begin{figure*}
\centering
\includegraphics[width= 1.0 \textwidth]{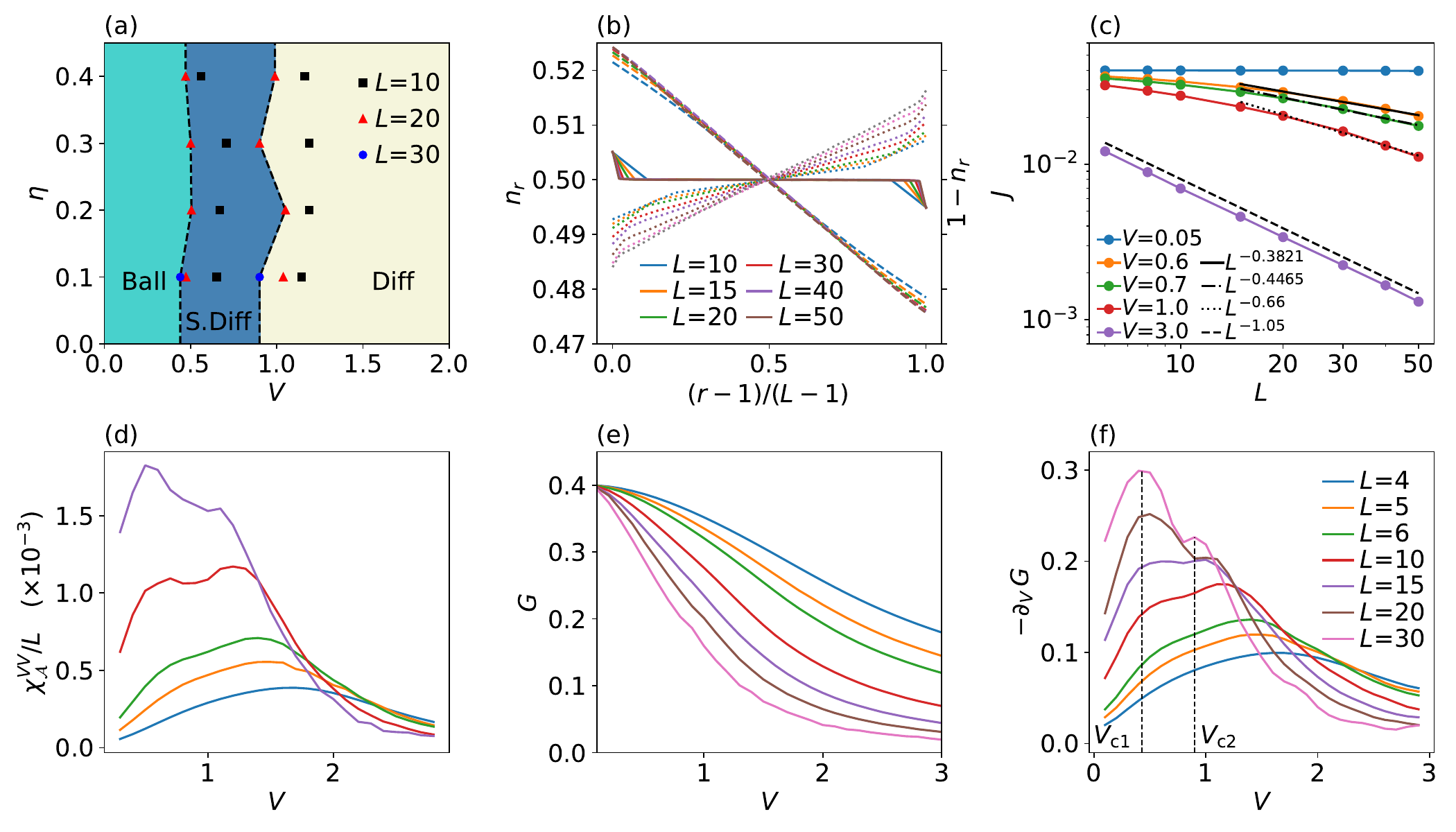}
\caption{\textbf{Results for the fermionic ladder.} (a) Tentative finite-sized phase diagram, showing a ballistic (Ball), diffusive (Diff) and super-diffusive (S. Diff) regimes. Phase boundaries lines are based on the largest converged system size. The symbols give the location of the peaks in (f) for different system sizes showcasing the drift of $V_\text{c1}$ and $V_\text{c2}$. (b) Comparison of the real space occupations between the ballistic phase at $V=0.05$ (continuous), the diffusive at $V=3.0$ (dashed) and the super-diffusive at $V=0.7$ (dotted); all for $\eta=0.05$. For the super-diffusive case (dotted) we showed $1-n_r$ to avoid overlapping the curves. (c) Comparison of the $L$ dependence of the current for the different regimes, with $\eta=0.05$. The straight lines serve as guides to the eye for the power-law behavior of the diffusive and super-diffusive phases. (d) $\chi^{VV}_{\mathcal{A}}/L$ vs $V$ for different system sizes at $\eta=0.1$ and with $\delta V=0.2$, detecting a transition from a ballistic to a diffusive phase, while crossing an intermediate super-diffusive regime.
The color code for figures (d)-(f) is given in (f).
(e) and (f) showcase $G$ and $-\partial_V  G$ (finite differences) vs $V$ for different $L$ and $\eta=0.1$, with the result in (f) smoothed with a low-pass filter to remove high-frequency noise.}
\label{fig.ladder}
\end{figure*}
%%%%%%%%%%%%%%%%%%%%%%%%%%%%%%%%%%%%%%%%%%%%%%%%%%%%%%%%%%

\subsection*{Fermionic ladder}

The second case we consider is that of an interacting two-leg ladder whose NESS behavior has so far not been addressed.  A sketch of the model is shown in Fig.\ref{sketches}-(b).
The system is described by Eq. \eqref{Ham}, with $i=(r,\tau)$, where $r=1,...,L$ labels the rungs and $\tau=1,2$ the legs. Hopping and interactions are restricted to nearest-neighbors of the ladder geometry. The jump operators, located at the ends of the ladder, are given by Eq.~\eqref{Jump} with $l=(r=1,L;\tau=1,2)$, $\eta_{(r=1,\tau=1,2)} = - \eta_{(r=L,\tau=1,2)} = \eta$  and $\Gamma_{(r=1,L,\tau=1,2)} = 1$. The full Hamiltonian governing the dynamics of this model is provided in \eqref{H_ladder} of the method section. 

Related systems featuring spins on a ladder were studied by \v{Z}nidari\v{c} \cite{Marko6,Marko7}.
The Hubbard model which can also be seen as a ladder system with the spin projection as the leg index has been widely studied \cite{Prosen5,Prosen6,Popkov2015}. 
It was found that, in contrast to the chain, the ballistic regime appears to be absent for locally coupled reservoirs under symmetric driving. 
Generic diffusive behaviour was also found for closed spin ladders in the linear-response regime using quantum typicality arguments \cite{Steinigeweg2014}.

 Motivated by the results of Refs. \cite{Prosen5,Marko6,Marko7} one may expect that the phase diagram of the fermionic ladder would feature only diffusive behavior. This expectation, however, is in contrast to the behavior of the $\chi_{\mathcal{A}}^{VV}$ vs. $V$, shown in Fig.\ref{fig.ladder}-(d) for different $L$. As $L$ increases, this quantity develops a two-peak structure hinting at the existence of two phase transitions, located at $V_{c1}$ and $V_\text{c2}$, and a much richer phase diagram.  

To further demonstrate the predictive power of  $\bm \Sigma$ we turn to a discussion of the properties of the different regimes.
In Fig.\ref{fig.ladder}-(c), we show $J$ as a function of $L$. 
Surprisingly, in addition to the ballistic and diffusive phases for small and large $V$ respectively, we find  a super-diffusive regime at intermediate $V$ for which $J\propto L^{-\nu}$, with $0<\nu<1$.  

Fig.\ref{fig.ladder}-(b) shows the density profiles for the three phases in one of the legs of the ladder. The profiles in the ballistic and diffusive regimes have a behaviour which is reminiscent of the corresponding phases of the chain. 
For the super-diffusive phase, the density profile also seems to depend linearly in $r$ near the middle of the system.

Fig.\ref{fig.ladder}-(a) depicts the phase diagram calculated from the maxima of $-\partial_V G$, shown in Fig.\ref{fig.ladder}-(f), computed from the conductance in  Fig.\ref{fig.ladder}-(e). 
We note that, although our results point to the existence of three distinct phases, we observe a small drift of the critical value of $V_\text{c1}$ and  $V_\text{c2}$ with system size. At the present stage, we thus cannot completely rule out that in the infinite system limit the $V_\text{c1}\to 0$, i.e. ballistic transport only arises at $V=0$, or even a more dramatic scenario where both $V_\text{c1}, V_\text{c2}\to 0$ and only the diffusive phase is stable for finite $V$. Although unlikely, disproving these scenarios will require further studies.  Either way, what is important in the present context is that $\chi_{\mathcal{A}}^{VV}$ is able to detect finite-size signatures of phase transitions for comparatively small system sizes.

\section*{Discussion}

We proposed the affinity susceptibility, and other measures of distance between the single particle correlations matrix, as multipurpose detectors for phase transitions where the order parameter is elusive. 
This situation occurs in certain systems in equilibrium but is prevalent in open  systems far from equilibrium.  

In contrast to the well known fidelity susceptibility, the affinity has the advantage of being available at a lower computational cost for commonly employed numerical methods, such as Monte-Carlo and MPS techniques. 

We demonstrated the usefulness and predictive power of the affinity susceptibility in two models of fermionic quantum matter out of equilibrium, sketched in Figs.\ref{sketches}-(a) and (b). The well known boundary driven fermionic chain of Fig.\ref{sketches}-(a) was used as a benchmark for the proposed method while the second model allowed us to test its predictive power.

For the first model, we recovered the known phase diagram and showed that the affinity susceptibility is enhanced at the phase transition already for relatively small sizes (see Figs.\ref{chain_pd}-(d) and (e)). 
For the boundary driven fermionic ladder, we used the affinity to uncover a non-trivial phase diagram. The two peak structure of the affinity susceptibility suggests the existence of two phase transitions upon increasing the interaction strength, see Fig.\ref{fig.ladder}-(d). This is an unexpected and surprising result as all existing results for spin ladders only feature diffusive behavior\cite{Prosen5,Marko6,Marko7}.

Motivated by these results, we preformed a thorough study of the finite size scaling of the current and the spacial density profile. This corroborated the predictions of the affinity, for all numerically accessible system sizes. 
Our data point to the existence of two phase transitions in the thermodynamic limit, where upon increasing the interaction the systems passes from ballistic to super-diffusive and subsequently to diffusive behavior. 

Another remarkable feature of the affinity susceptibility is that in all the cases we studied it signals phase transitions already for comparatively small system sizes. This suggest that corrections to scaling of the affinity susceptibility are smaller than those for other quantities. Yet, perhaps because it also encodes long-range correlations, a higher accuracy is required in the convergence of the MPO as compared to local operators like current and density. For this reason only relatively smaller sizes were considered in the case of the ladder. 
    
These findings underline the potential of the affinity susceptibility as an indicator for phase changes in open systems.

It will be interesting to exploit  the affinity to study other instances where single-particle correlators are the only easily available quantities. This applies to Quantum Monte Carlo studies out of equilibrium or at finite temperatures and in more than one dimension. Our method might also prove useful for studying time dependent critical phenomena away from equilibrium where effective techniques are badly needed.

\section*{Methods} \label{sec:Method}

\subsection*{Explicit Hamiltonians for both models}

To avoid large cumbersome expressions, we presented in Eq.\eqref{Ham} a compressed form of the Hamiltonian and provided sketches illustrating the details of the interactions in Fig.\ref{sketches}-(a) and (b). For the sake of completeness, the explicit expressions for the Hamiltonian for the fermion chain model is given by
\begin{align}
H= & -t \sum_{r=1}^{L-1} \left[ c^\dagger_r \, c_{r+1} + c^\dagger_{r+1} \, c_{r} \right] \nonumber \\ 
& + V \sum_{r=1}^{L-1} \left( c^\dagger_r \, c_r - \frac{1}{2} \right) \left( c^\dagger_{r+1} \, c_{r+1} - \frac{1}{2} \right) \quad,
\label{H_chain}
\end{align}
while for the ladder
\begin{align}
H= & -t \sum_{r=1}^{L-1} \sum_{\tau=1,2} \left[ c^\dagger_{r,\tau} \, c_{r+1,\tau} + c^\dagger_{r+1,\tau} \, c_{r,\tau} \right] \nonumber \\
&  -t \sum_{r=1}^L \left[ c^\dagger_{r,1} \, c_{r,2} + c^\dagger_{r,2} \, c_{r,1} \right] \nonumber \\ 
& + V \sum_{r=1}^{L-1} \sum_{\tau=1,2} \left( c^\dagger_{r,\tau} \, c_{r,\tau} - \frac{1}{2} \right) \left( c^\dagger_{r+1,\tau} \, c_{r+1,\tau} - \frac{1}{2} \right) \nonumber \\
& + V \sum_{r=1}^L \left( c^\dagger_{r,1} \, c_{r,1} - \frac{1}{2} \right) \left( c^\dagger_{r,2} \, c_{r,2} - \frac{1}{2} \right) \quad ,
\label{H_ladder}
\end{align}
where $\tau=1,2$ labels the legs of the ladder.

\subsection*{Properties of the single-particle affinity} 

In this section we present details regarding how the single-particle affinity is numerically evaluated and discuss additional properties. In particular, we demonstrate that the affinity coincides with the mixed-state fidelity for the case of quadratic systems. This allows us to show that it satisfies the required properties of a distance between covariance matrices. 
Finally, we also show that the results in the main text 
are expected to hold for other notions of distance between covariance matrices.

\subsubsection*{Details about the numerical evaluation}

The affinity susceptibility is defined in terms of the expansion of the affinity, shown in Eq.\eqref{FidX}. Defining $\mathcal A(\delta) = \mathcal A[\boldsymbol \Sigma (V), \boldsymbol \Sigma (V+\delta) ] $, we can write
\cite{Rossini2018,Mera2018}
\begin{align}
\chi^{VV}_{\mathcal{A}}(\delta) &  = - \lim_{\delta \to 0} \frac{\mathcal A(\delta)+ \mathcal A(-\delta) - 2\mathcal A(0)}{\delta^2}.
\end{align}
We evaluate this quantity numerically by fixing the value of the perturbation $\delta$ to  be small and consider states at points $V$, $V+\delta$ and $V-\delta$. Note that, since $\Sigma$ is obtained at  finite numerical precision,  the value of $\delta$ has to be taken sufficiently large to ensure that the difference between the two covariance matrices is much larger than that precision.

\subsubsection*{Relation with fidelity for free fermions}

For quadratic systems the density operator is given by a Gaussian
\begin{equation}
\rho = \frac{1}{Z} e^{-\frac{1}{2} \mathbf{C}^\dagger \, \boldsymbol{\Omega} \, \mathbf{C}} \quad ,
\label{rho_free}
\end{equation}
where $\mathbf{C}= \left(c_1,\cdots,c_L,c_1^\dagger,\cdots,c_L^\dagger \right)^T$ and $\boldsymbol{\Omega}$ is a Hermitian matrix with the particle-hole symmetric structure 
\begin{equation}
\boldsymbol{\Omega} = 
\begin{pmatrix}
\mathbf{h} & \boldsymbol{\Delta} \\
\boldsymbol{\Delta}^\dagger & -\mathbf{h}^T 
\end{pmatrix} \quad .
\end{equation}
Each of its blocks is a $L\times L$ matrix, with $\mathbf{h}=\mathbf{h}^\dagger$ and $\boldsymbol{\Delta}^T=-\boldsymbol{\Delta}$. For thermal states $\boldsymbol{\Omega}$ can be seen as the Hamiltonian divided by temperature. $Z$ is the partition function, which for quadratic models can be written 
\begin{align}
Z &= \Tr \left[ e^{-\frac{1}{2} \mathbf{C}^\dagger \, \boldsymbol{\Omega} \, \mathbf{C}} \right]  = \sqrt{\det \left[ \mathbf{1} + e^{-\boldsymbol{\Omega}} \right] } \quad .
\label{Z_quad}
\end{align}

The covariance matrix also acquires a simple form:
\begin{align}
\bm{\Sigma} &= \frac{1}{Z} \Tr \left[ e^{-\frac{1}{2} \mathbf{C}^\dagger \, \boldsymbol{\Omega} \, \mathbf{C}} \, \mathbf{C} \, \mathbf{C}^\dagger \right] = \left[ \mathbf{1} + e^{-\boldsymbol{\Omega}} \right]^{-1} \quad .
\label{X_quad}
\end{align}

From the definition of the fidelity 
\begin{equation}
F(\rho_1,\rho_2) = \left( \Tr \, \sqrt{\sqrt{\rho_1} \, \rho_2 \, \sqrt{\rho_1} } \right)^2 \quad ,
\label{Fid_def}
\end{equation}
and using the identity 
\begin{equation}
e^{-\frac{1}{2} \mathbf{C}^\dagger \, \boldsymbol{\Omega}_1 \, \mathbf{C}} \, e^{-\frac{1}{2} \mathbf{C}^\dagger \, \boldsymbol{\Omega}_2 \, \mathbf{C}} = e^{-\frac{1}{2} \mathbf{C}^\dagger \, \tilde{\boldsymbol{\Omega}} \, \mathbf{C}} \quad ,
\end{equation}
where $\tilde{\boldsymbol{\Omega}}$ is defined as $e^{- \tilde{\boldsymbol{\Omega}} } = e^{- \boldsymbol{\Omega}_1 } \, e^{- \boldsymbol{\Omega}_2 }$, 
we can write
\begin{align}
F(\rho_1,\rho_2) &= \frac{1}{Z_1 \, Z_2} \left( \Tr \, \sqrt{ e^{-\frac{1}{2} \mathbf{C}^\dagger \, \boldsymbol{\Omega}_1 \, \mathbf{C}} \, e^{-\frac{1}{2} \mathbf{C}^\dagger \, \boldsymbol{\Omega}_2 \, \mathbf{C}}} \right)^2 \nonumber \\
&= \frac{1}{Z_1 \, Z_2}  \det \left[ \mathbf{1} + \sqrt{e^{-\boldsymbol{\Omega}_1} \, e^{-\boldsymbol{\Omega}_2} } \right] \nonumber \\
&= \frac{ \det \left[ \mathbf{1} + \sqrt{\left( \bm{\Sigma}_1^{-1} - \mathbf{1} \right) \left( \bm{\Sigma}_2^{-1} - \mathbf{1} \right) } \right] }{\sqrt{\det \left[ \bm{\Sigma}_1^{-1} \right] \det \left[ \bm{\Sigma}_2^{-1} \right] }} 
\label{Fid_aff}
\end{align}
where $\bm{\Sigma}_i$  is the covariance matrix corresponding to $\rho_i$. 

Note that in Eq. \eqref{Fid_aff} we assumed that we can invert the covariance matrices. 
Nevertheless, the expression still has a well defined value in the limit where $\bm{\Sigma}_i$ is not invertible.

\subsubsection*{Notion of distance for the single-particle affinity}

A metric is a function $D: X \times X \to [0,\infty[$ that provides a distance between two members of some set $X$. It has to obey the following properties for all $x,y,z \in X$\cite{Ingemar2017}:
\begin{itemize}
\item{\makebox[4.0cm][l]{$D(x,y)=0$ $\Leftrightarrow$ $x=y$} identity of indiscernibles}
\item{\makebox[4.0cm][l]{$D(x,y)=D(y,x)$}  symmetry}
\item{\makebox[4.5cm][l]{$D(x,y)\leq D(x,z)+D(z,y)$} triangle inequality.}
\end{itemize}

The fidelity does not actually constitute a metric between density operators, but it can be related to other quantities that do: the Bures distance $D_{\mathcal B}(\rho_1,\rho_2) =\sqrt{ 2 - 2\sqrt{F (\rho_1,\rho_2) }}$ \cite{Bures69,Uhlmann76,Jozsa94} and the Bures angle $D_\alpha=\arccos \sqrt{F(\rho_1,\rho_2)}$ \cite{Uhlmann95}. In the previous section we showed how the single-particle affinity corresponds to the fidelity for quadratic systems. By continuation, it follows that it can be related to the notion of distance between states.

\subsubsection*{Other notions of distance}

So far we focused on the single-particle affinity and its susceptibility to detect out-of-equilibrium phase transitions. 
To emphasize that the quantity that contains the relevant information  is the covariance matrix, we discuss here a similar analysis using different notions of distance and verify that the results agree qualitatively.   

Consider the Bhattacharyya distance \cite{Bhattacharyya}  between classical probability distributions $D_{\text{Bhatt}}(p,q)= - \log \int \sqrt{p(x) \, q(x)} \, dx$. 
When evaluated on Gaussian distributions with zero mean, the Bhattacharyya distance writes 
\begin{equation}
D_{\text{Bhatt}} (p,q)=   \frac{1}{2} \log \left( \frac{\det \frac{C_q+C_p}{2}}{\sqrt{\det C_q \, \det C_p}} \right),
\end{equation}
in terms of the covariance matrices $C_p$ and $C_q$. This expression induces a distance between real covariant matrices. 
Its generalisation to Hermitian matrices yields   
\begin{equation}
D_{\text{Bhatt}} (\bm{\Sigma}_1,\bm{\Sigma}_2)=   \frac{1}{2} \log \left( \frac{\det \frac{\bm{\Sigma}_1+\bm{\Sigma}_2}{2}}{\sqrt{\det \bm{\Sigma}_1 \, \det \bm{\Sigma}_2}} \right).
\end{equation}

In the same way, from the Wasserstein distance \cite{Wasserstein} we obtain 
\begin{equation}
D_{\text{Wass}} (\bm{\Sigma}_1,\bm{\Sigma}_2)= \tr \left[ \bm{\Sigma}_1 + \bm{\Sigma}_2 - 2 \sqrt{ \sqrt{\bm{\Sigma}_2} \, \bm{\Sigma}_1 \, \sqrt{\bm{\Sigma}_2}}\right] .
\end{equation}

Finally, we also consider the Trace distance \cite{Trace}
\begin{equation}
D_{\text{trace}} (\bm{\Sigma}_1,\bm{\Sigma}_2)= \frac{1}{2} \tr \, |\bm{\Sigma}_1-\bm{\Sigma}_2| ,
\end{equation}
where $|A|=\sqrt{A^\dagger \, A}$.

In the supplementary material--S1 we illustrate these different distances in the two examples considered before.

\subsection*{Open systems with MPS}

The vectorized form of generic density matrix $\rho_{\bm{\sigma},\bm{\sigma}'}$, with  $\bm{\sigma} = \sigma_1 \sigma_2 \hdots \sigma_L$ is given by 
\begin{equation}
\ket{\rho}\rangle = \sum_{\bm{\sigma},\bm{\sigma}'} \rho_{\bm{\sigma},\bm{\sigma}'} \ket{\bm{\sigma},\bm{\sigma}'}\rangle,
\label{rho_MPO}
\end{equation}
with $\ket{\bm{\sigma},\bm{\sigma}'}\rangle=\ket{\bm{\sigma}} \bra{\bm{\sigma}'}^T $.
It admits the MPO decomposition 
\begin{equation}
 \ket{\rho}\rangle = \sum_{\mathbf{s}} M^{s_1}\hdots M^{s_i} \hdots M^{s_L} \ket{\mathbf{s}}.
\end{equation}
where $s_i=(\sigma_i,\sigma_i')$, and $M^{\sigma_i,\sigma'_i}$ are the MPO decomposition matrices with dimension up to the bond dimension $D$. 
This yields a MPS with a local dimension encompassing the original Hilbert space and that of the copy. Using the vector notation, the integrated Lindblad dynamics of Eq. \eqref{Lind} is
\begin{equation}
\ket{\rho(t)}\rangle = e^{\hat{\mathcal{L}} t} \ket{\rho_0}\rangle .
\end{equation}
In this form the existing time-evolution algorithms for MPS and unitary dynamics \cite{Vidal2,Daley} can readily be applied here without significant modifications \cite{Vidal3,Cirar1}.

For this work we used the t-DMRG algorithm for time-evolution with a Trotter decomposition of 4th order as described in Ref. \cite{Prosen8} and with an associated error per iteration of $O(\Delta t^4)$, where $\Delta t$ is the time step. 

We used the ITensor library \cite{itensor} as the basis of our implementation. In the initial stages of the evolution a larger time step was chosen, typically in the range $\Delta t\in [0.1,0.5]$, to speed up convergence. In the final stages, when necessary, we switched to a smaller time step to better approximate the steady-state, but generally not smaller than $\Delta t=0.01$.

To guarantee the correctness of the results, the following recipe was used \cite{Ferreira2}.  
For a set of parameters $L$, $V$, $\eta$ and $D$,
\begin{itemize}
\item we monitored $J$ in the middle bond during time-evolution until it saturated. The condition for convergence was $\sigma_t/J_t<0.01$, where $J_t$ and $\sigma_t$ are respectively the mean value and the standard deviation of the last 50 values of $J$ obtained during time-evolution for the middle bond. This ensured that fluctuations only affected digits at least 2 decimal places after the most significant one;

\item the obtained steady-states are supposed to possess a constant current across the length of the system, which was tracked by checking if $\sigma_x/J_x<0.01$, where $J_x$ and $\sigma_x$ are the mean value and standard deviation for the current at the different bonds. If this condition was not fulfilled, the state was evolved further in time until it was;

\item to determine if the MPO description approximates sufficiently well a given steady-state, we also analyzed the convergence with bond dimension. The criterion for convergence of the current with bond dimension, which was applied for prototypical cases, was $\sigma_D/J_D<0.01$, where $J_D$ and $\sigma_D$ are respectively the mean value and standard deviation of a set composed of $J_x$, for the largest bond dimensions used. Typically, the bond dimension of the results shown in the main text was $D=100$, but it went up to $D=150$ for the larger system sizes.
\end{itemize}

\section*{Acknowledgements}

We gratefully acknowledge Cheng Chen for useful discussions regarding the MPO implementation.
Computations were performed on the Tianhe-2JK cluster at the Beijing Computational Science Research Center (CSRC), on the Baltasar-Sete-S\'ois cluster, supported by V. Cardoso's H2020 ERC Consolidator Grant No. MaGRaTh-646597, computer assistance was provided by CSRC and CENTRA/IST; and on the OBLIVION Supercomputer (based at the High Performance Computing Center - University of \'Evora) funded by the ENGAGE SKA Research Infrastructure (reference POCI-01-0145-FEDER-022217 - COMPETE 2020 and the Foundation for Science and Technology, Portugal) and by the BigData@UE project (reference ALT20-03-0246-FEDER-000033 - FEDER and the Alentejo 2020 Regional Operational Program).
M. M. Oliveira acknowledges support by FCT through Grant No. SFRH/BD/137446/2018. PR acknowledge support by FCT through Grant No. UID/CTM/04540/2019 and by the QuantERA II Programme that has received funding from the European Union’s Horizon 2020 research and innovation programme under Grant Agreement No 101017733.
S.\,K.\, acknowledges support by the  by National Science and Technology Council, Taiwan (grant No. NSTC 111-2634-F-A49-007), the Featured Area Research Center Program within the framework of the Higher Education Sprout Project by the Ministry of Education (MOE) in Taiwan and the Yushan Fellowship Program of the  MOE  Taiwan.

\bibliographystyle{apsrev4-1}
\bibliography{References}

\pagebreak 
\newpage

\begin{widetext}
\begin{center}
\textbf{\large{}\textemdash{} Supplemental Material \textemdash{}}
\par\end{center}{\large \par}
\begin{center}
\textbf{\large{}Efficient quantum information probes of non-equilibrium quantum criticality}
\par\end{center}{\large \par}
\begin{center}
\textbf{Miguel M. Oliveira$^{(a)}$, Pedro Ribeiro$^{(a,b)}$, and Stefan Kirchner$^{(c,d)}$}\\
$^{(a)}$CeFEMA, Instituto Superior T\'ecnico, Universidade de Lisboa Av. Rovisco
Pais, 1049-001 Lisboa, Portugal\\
$^{(b)}$Beijing Computational Science Research Center, Beijing 100193, China\\
$^{(c)}$Department of Electrophysics, National Yang Ming Chiao Tung University, Hsinchu 30010, Taiwan\\
$^{(d)}$Center for Emergent Functional Matter Science, National Yang Ming Chiao Tung University, Hsinchu 30010, Taiwan
\end{center}
\begin{description}
\item [{Summary}] Below we provide additional technical details and further numerical results supplementing the conclusions from the main text.
\end{description}
\end{widetext}

\beginsupplement

\section{Other notions of distance}

In the methods section of the main text we introduced different notions of distance between covariance matrices, namely the Bhattacharyya, Wasserstein and Trace distances. This was done to emphasize that the quantity that contains the relevant information is the covariance matrix; and that the analysis in the main text can be replicated with these alternative distances. 

\begin{figure}[t!]
\centering
\includegraphics[width= 0.5 \textwidth]{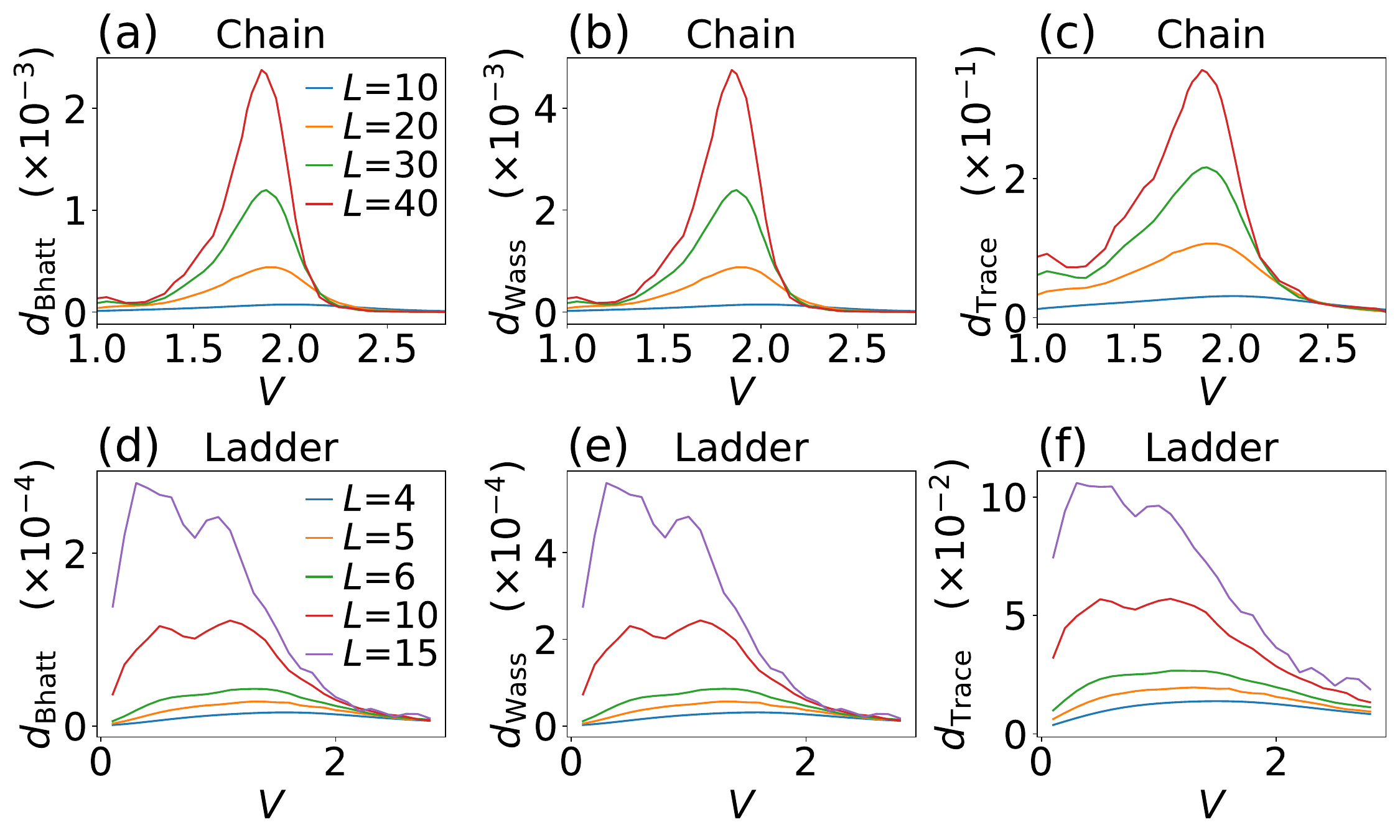} 
\caption{\textbf{Alternative notions of distance as a function of $V$ for different system sizes.} (a), (b) and (c) correspond respectively to the Bhattacharyya, Wasserstein and Trace distances for the chain model. (d), (e) and (f) show the same quantities for the ladder model.}
\label{dist_alt}
\end{figure}

To demonstrate this equivalence, in Figs.\ref{dist_alt}-(a), (b) and (c) we show these different distances as a function of $V$ for the chain model. Figs.\ref{dist_alt}-(d), (e) and (f) depict the same for the ladder case. Both exhibit compatible behavior with the results shown in Fig.\ref{chain_pd}-(e) and \ref{fig.ladder}-(d) of the main text. 

Note that here the distances are computed using only two points, taken at $V$ and $V+\delta V$, which had the effect of slightly left-shifting the results with respect to what was shown in the main text.

%%%%%%%%%%%%%%%%%%%%%%% Figure 1 of Appendix 1 %%%%%%%%%%%%%%%%%%%%%%%%%%
%%%%%%%%%%%%%%%%%%%%%%%%%%%%%%%%%%%%%%%%%%%%%%%%%%%%%%%%%%

\section{$\eta=1$ Critical region}

The $\eta=1$ states correspond to a special configuration of the reservoirs where electrons are only injected on the left and removed from the right of the system. As mentioned in the main text, the $\eta=1$ line of the phase diagram of both models corresponds to a critical region, which is signaled by a peak in the SPA susceptibility $\chi^{\eta \eta}_{\mathcal{A}}$. This not only happens for $V>V_\text{c}$, where the system transitions from a diffusive to an insulating regime as $\eta$ approaches 1; but also for $V<V_\text{c}$ inside the ballistic regime. 

\begin{figure}[t!]
\centering
\includegraphics[width= 0.5 \textwidth]{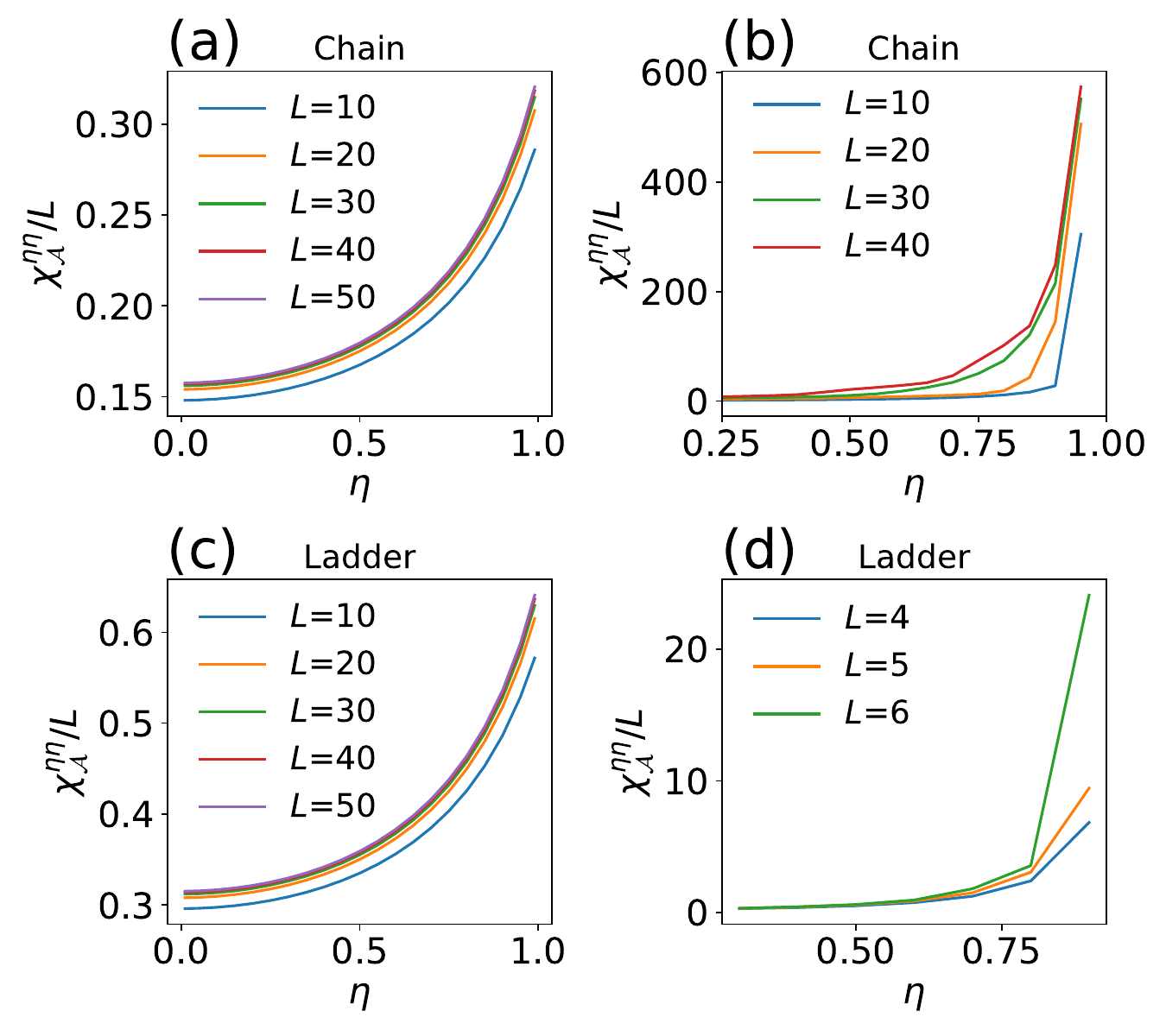} 
\caption{\textbf{Affinity susceptibility as a function of $\eta$ for different system sizes.} (a) for the chain at $V=0$; (b) for the chain at $V=4$; (c) for the ladder at $V=0$; (d) for the ladder at $V=2$. The results in (a) and (c) were obtained with ED, which for the non-interacting case is numerically efficient, with a perturbation of $\delta \eta=0.01$. MPS methods were used to obtain (b) and (d), for which a perturbation of $\delta \eta=0.05$ and $\delta \eta=0.1$ were used respectively.}
\label{critical_mu}
\end{figure}

In Fig.\ref{critical_mu}-(a) and (b) we show the SPA susceptibility $\chi^{\eta \eta}_{\mathcal{A}}$ per degree of freedom for the chain model at $V=0$ and $V=4$ respectively. In Figs.\ref{critical_mu}-(c) and (d) we show the same quantity for the ladder at $V=0$ and $V=2$. In all of the figures $\chi^{\eta \eta}_{\mathcal{A}}$ diverges as $\eta$ approaches 1. The insulating regime is hard to converge \cite{Benenti2}, which explains the comparatively small system sizes for the ladder in Fig.\ref{critical_mu}-(d).

\begin{figure}[t!]
\centering
\includegraphics[width= 0.5 \textwidth]{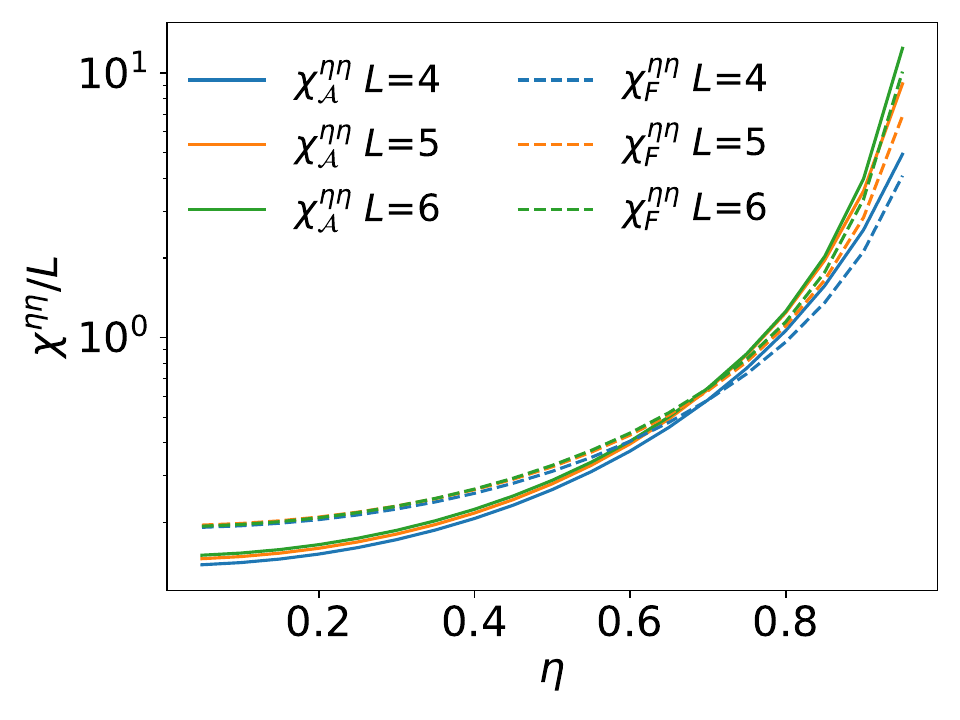} 
\caption{\textbf{Comparison between affinity susceptibility (full) and fidelity susceptibility (dashed) as a function of $\eta$ for different system sizes.} Both quantities exhibit compatible behavior. The perturbation used was $\delta \eta=0.01$ and the results were obtained with ED for the fermion chain model.}
\label{critical_mu_fid}
\end{figure}

Similarly to what was done in Fig.\ref{chain_pd}-(d) of the main text, where a comparison between the affinity susceptibility $\chi^{VV}_{\mathcal{A}}$ and the fidelity susceptibility $\chi^{VV}_{F}$ as a function of $V$ was shown; in Fig.\ref{critical_mu_fid} a comparison of $\chi^{\eta\eta}_\mathcal{A}$ and $\chi^{\eta\eta}_F$ is shown for the chain model. Evidently, both quantities exhibit compatible behavior.

%%%%%%%%%%%%%%%%%%%%%%% Figure 1 of Appendix 3 %%%%%%%%%%%%%%%%%%%%%%%%%%
%%%%%%%%%%%%%%%%%%%%%%%%%%%%%%%%%%%%%%%%%%%%%%%%%%%%%%%%%%

%%%%%%%%%%%%%%%%%%%%%%% Figure 2 of Appendix 3 %%%%%%%%%%%%%%%%%%%%%%%%%%
%%%%%%%%%%%%%%%%%%%%%%%%%%%%%%%%%%%%%%%%%%%%%%%%%%%%%%%%%%

\section{Order Parameters} 

The distinct $L$-dependence of the ballistic and diffusive regimes prompted us to propose quantities that can serve as potential order parameters. 

The first quantity was $G$, which we dubbed conductance by way of analogy with Ohm's law. This was defined as $G=J/\eta$, where $J$ is the current and $\eta$ the injection/removal imbalance of the Markovian reservoirs. $G$ is finite in the ballistic regime, but goes to zero in the thermodynamic limit for the diffusive phase, see Fig.\ref{chain_pd}-(c) of the main text.

\begin{figure}[t!]
\centering
\includegraphics[width= 0.5 \textwidth]{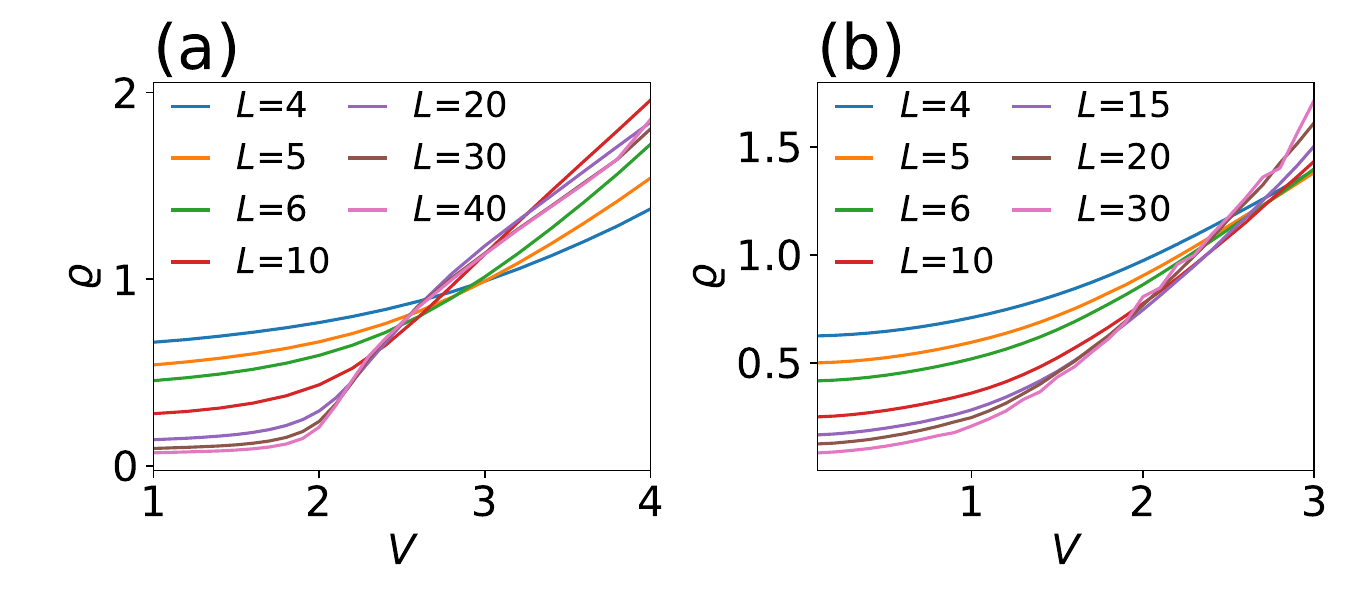} 
\caption{\textbf{Resistivity $\varrho$ as a function of $V$.} (a) for the chain and (b) for the ladder.}
\label{resistivity}
\end{figure}

Similar considerations apply to the resistivity $\varrho$. We define it in terms of the induced resistance $R = \varrho L = 1/G$, such that $\varrho = \eta/(J L)$. In contrast to the conductance, $\varrho$ is finite in the diffusive phase and goes to zero on the ballistic regime, see Fig.\ref{resistivity}-(a).

So far we focused on the chain model; for the ladder one reads off Fig.\ref{fig.ladder}-(e) of the main text that $G$ acquires a $L$ dependence for small interactions, which  increases with increasing $V$. This is compatible with the hypothesis of a ballistic phase that transitions into a diffusive regime by crossing a super diffusive region. 
One might speculate that $G$ and $\varrho$ can be used to extract 
estimates for $V_\text{c1}$ and $V_\text{c2}$. In Fig.\ref{resistivity}-(b), we can see for the fermion ladder model $\varrho$ as function of $V$ at different system sizes. Its behavior is complementary with that of the conductance, but we were not able to use it to estimate the position of the critical points. In the main text we used instead the derivative of the conductance for that purpose, see Fig.\ref{fig.ladder}-(f).

%%%%%%%%%%%%%%%%%%%%%%% Figure 1 of Appendix 4 %%%%%%%%%%%%%%%%%%%%%%%%%%
%%%%%%%%%%%%%%%%%%%%%%%%%%%%%%%%%%%%%%%%%%%%%%%%%%%%%%%%%%

\section{Jordan-Wigner Transformation for Open Systems} 

The MPS formalism is more naturally implemented in terms of spin degrees of freedom. Its application to fermionic systems  is accomplished via the Jordan-Wigner (JW) transformation  \cite{Jordan1928,Lieb1961}. For short-range one-dimensional models the Hamiltonian remains local after the transformation and thus amenable to MPS techniques. 

The JW transformation is defined as
\begin{align}
&c_i = S_i \sigma^-_i \,  \nonumber \\
&c^\dagger_i =  S_i \, \sigma^+_i 
\label{JW-transf}
\end{align}
where $c_i$ ($c^\dagger_i$) are fermionic annihilation (creation) operators, $\sigma^\pm=(\sigma^z\pm i \, \sigma^y)/2$ are the spin-$\frac{1}{2}$ ladder operators,  $\sigma^{x,y,z}$ are the Pauli matrices, and $S_i=\left[ \prod_{j=1}^{i-1} \left( -\sigma^z_j \right) \right]$ is the string operator.

\subsection*{Chain Model}

The JW transformation requires an ordering of the sites. For a closed one-dimensional system, a sequential ordering yields a local Hamiltonian. 

For the chain this yields 
\begin{equation}
H= -\frac{t}{2} \sum_{i=1}^{L-1} \left[ \sigma^x_i \, \sigma^x_{i+1} + \sigma^y_i \, \sigma^y_{i+1} \right] + \frac{V}{4} \sum_{i=1}^{L-1} \sigma^z_i \, \sigma^z_{i+1} .
\end{equation}
and
\begin{align}
W_{1,\pm } & = \sqrt{\Gamma \frac{1\pm \eta}{2}} \sigma^\pm_1 \nonumber \\
W_{L,\pm } & = \mp \sqrt{\Gamma \frac{1 \mp \eta}{2}} (-1)^{\mathcal{\hat{N}}} \sigma^\pm_L
\end{align}
where we used in the expression for $W_{L,\pm }$  that  $S_{L}\sigma^\pm_L = \mp (-1)^{\mathcal{\hat{N}}} \sigma^\pm_L$ where 
$\hat{\mathcal{N}}$ counts the number of occupied fermion (or spin up) states.

%Since the Lindblad superoperator, shown in Eq.\eqref{Lindblad}, does not mix sectors with $(-1)^{\mathcal{\hat{N}}}=\pm1$, this quantity is a scalar within each sector (\pr{I THINK} this is not true ). Here, we only consider $(-1)^{\mathcal{\hat{N}}}=1$, since this is the only physical sector for the initial electronic model. 

We now have to consider the effect of the $(-1)^{\mathcal{\hat{N}}}$ operator on the Lindblad equation, see Eq.\eqref{Lindblad} of the main text. The only non-trivial term is the one where the jump operator acts from both the left and the right. 
To deal with this it is helpful to consider that the density matrix has two sectors that are not mixed by the Lindblad equation, the $\{\ket{\text{even}} \bra{\text{even}} , \ket{\text{odd}} \bra{\text{odd}} \}$ and $\{\ket{\text{even}} \bra{\text{odd}} , \ket{\text{odd}} \bra{\text{even}} \}$. The operator $(-1)^{\mathcal{\hat{N}}}$ commutes with the former and anti-commutes with the later, so we can replace it with $1$ and $-1$ respectively. 
Expectation values of operators composed of an even number of creation and annihilation operators only have contributions from the sector $(-1)^{\mathcal{N}}=1$; therefore this sector alone was considered.

\subsection*{Ladder Model}

For the fermion ladder model, regardless of the choice of ordering, the string operator always leads to the appearance of additional interaction terms in the spin model. 

\begin{figure}[t!]
\centering
\includegraphics[width= 0.2 \textwidth]{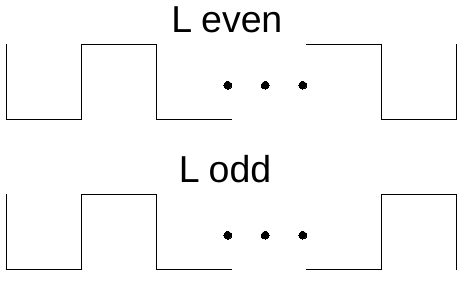} 
\caption{ \textbf{Jordan-Wigner ordering for the fermion ladder.}}
\label{JW_order}
\end{figure}

For the results in the main text, we chose the zig-zag ordering in Fig.\ref{JW_order}, which maintains the locality of the Hamiltonian. As suggested by the picture, nearest neighbor sites can have string operators in the intra-chain hopping terms, depending on whether the corresponding bond is even or odd.

%%%%%%%%%%%%%%%%%%%%%%% Figure 1 of Appendix 4 %%%%%%%%%%%%%%%%%%%%%%%%%%
%%%%%%%%%%%%%%%%%%%%%%%%%%%%%%%%%%%%%%%%%%%%%%%%%%%%%%%%%%

For clarity, instead of giving the transformed spin Hamiltonian and jump operators we list below the transformation rules.  

Labeling the rungs by $\tau=1,2$ for the upper and lower chain respectively, the intra-chain hopping terms of the Hamiltonian in Eq.\eqref{Ham} of the main text transform according to 
\begin{align}
&c^\dagger_{i,\tau} \, c_{i+1,\tau} \to \sigma^+_{i,\tau} \, \sigma^-_{i+1,\tau} \nonumber \\
&c^\dagger_{i+1,\tau} \, c_{i,\tau} \to \sigma^-_{i,\tau} \, \sigma^+_{i+1,\tau} \nonumber \\
&c^\dagger_{i,\bar{\tau}} \, c_{i+1,\bar{\tau}} \to \sigma^+_{i,\bar{\tau}} \, \sigma^z_{i,\tau} \, \sigma^z_{i+1,\tau} \, \sigma^-_{i+1,\bar{\tau}} \nonumber \\
&c^\dagger_{i+1,\bar{\tau}} \, c_{i,\bar{\tau}} \to \sigma^-_{i,\bar{\tau}} \, \sigma^z_{i,\tau} \, \sigma^z_{i+1,\tau} \, \sigma^+_{i+1,\bar{\tau}} \quad .
\end{align}
For an even bond we have that $\tau=1$ and $\bar{\tau}=2$, and for an odd bond $\tau=2$ and $\bar{\tau}=1$. For the inter-chain hopping and interaction terms the corresponding operators transform as
\begin{align}
&c^\dagger_{i,\tau} \, c_{i,\bar{\tau}} \to \sigma^+_{i,\tau} \, \sigma^-_{i,\bar{\tau}} \nonumber \\
&c^\dagger_{i,\tau} \, c_{i,\tau} - 1/2 \to \sigma^z_{i,\tau}/2 .
\end{align}
Using the same reasoning as for the chain model, we can obtain the transformation rules for the jump operators. For the left side of the ladder we get
\begin{align}
&c_{1,1} \to \sigma^-_{1,1} \nonumber \\
&c^\dagger_{1,1} \to \sigma^+_{1,1} \nonumber \\
&c_{1,2} \to  \sigma^z_{1,1} \, \sigma^-_{1,2} \nonumber \\
&c^\dagger_{1,2} \to  \sigma^z_{1,1} \, \sigma^+_{1,2} .
\end{align}
On the right side we obtain 
\begin{align}
&c_{L,\tau} \to \sigma^-_{L,\tau} \nonumber \\
&c^\dagger_{L,\tau} \to \sigma^+_{L,\tau} \nonumber \\
&c_{L,\bar{\tau}} \to  \sigma^z_{L,\tau} \, \sigma^-_{L,\bar{\tau}} \nonumber \\
&c^\dagger_{L,\bar{\tau}} \to  \sigma^z_{L,\tau} \, \tau^+_{L,\bar{\tau}} \quad ,
\end{align}
where for $L$ even we have $\tau=1$ and $\bar{\tau}=2$, and for $L$ odd $\tau=2$ and $\bar{\tau}=1$.

\subsection*{Observables}

In this subsection, it is  shown how to relate the spin and fermionic observables used to characterize the different NESS regimes. 

\subsubsection*{Occupation}

The local occupation transform according to
\begin{equation}
\langle n_i \rangle = \frac{\langle \sigma^z_i \rangle + 1}{2},
\end{equation}
which is valid for both the chain and ladder models.

\subsubsection*{Current}

The current operator in the chain model is written in terms of the spin degrees of freedom as
\begin{align}
J^{\text{chain}}_i & = -\frac{t}{2} \left \langle \sigma^x_i \, \sigma^y_{i+1} - \sigma^y_i \, \sigma^x_{i+1} \right \rangle .
\end{align}
For the ladder one finds
\begin{align}
J^{\text{ladder}}_i &= -\frac{t}{2} \left\langle \sigma^x_{i,\tau} \, \sigma^y_{i+1,\tau} - \sigma^y_{i,\tau} \, \sigma^x_{i+1,\tau} \right\rangle \nonumber \\
& -\frac{t}{2} \left\langle \sigma^x_{i,\bar{\tau}} \, \sigma^z_{i,\tau} \, \sigma^z_{i+1,\tau} \, \sigma^y_{i+1,\bar{\tau}} - \sigma^y_{i,\bar{\tau}} \, \sigma^z_{i,\tau} \, \sigma^z_{i+1,\tau} \, \sigma^x_{i+1,\bar{\tau}} \right\rangle,
\end{align} 
where $\tau=1$ and $\bar{\tau}=2$ for an even bond, and $\tau=2$ and $\bar{\tau}=1$ for an odd one.

\subsubsection*{Covariance Matrix}

Consider the block decomposition of the covariance matrix
\begin{equation}
\bm{\Sigma} =  \left(\begin{array}{cc}
\av{cc^\dagger} & \av{cc}\\
\av{c^\dagger  c^\dagger} & \av{c^\dagger c}
\end{array}\right) .
\end{equation}
The first two blocks are given by 
\begin{equation}
\av{c_i c^\dagger_j} = \left \langle \sigma^-_i \, S_{i,j} \, \sigma^+_j \right \rangle; \ \
\av{c_i c_j} = \left \langle \sigma^-_i \, S_{i,j} \, \sigma^-_j \right \rangle ,
\end{equation}
where we assumed, without loss of generality, that $i<j$ and $S_{i,j} = \left [ \prod_{k=i}^{j-1} (-\sigma^z_k) \right]$. 
The other cases can be obtained by taking advantage of the symmetries of the covariance matrix. The string operator for the chain will be a straight line between the sites $i$ and $j$, whereas for the ladder it will be a zig-zag line, see Fig.\ref{JW_order}.

\end{document}